# Two-step stabilization of orbital order and the dynamical frustration of spin in the model charge-transfer insulator KCuF$_3$

James C. T. Lee[1]*, Shi Yuan[1]*, Siddhartha Lal[1,2]*, Young Il Joe[1], Yu Gan[1], Serban Smadici[1], Ken Finkelstein[3], Yejun Feng[4], Andrivo Rusydi[5], Paul M. Goldbart[1,2], S. Lance Cooper[1], and Peter Abbamonte[1]

*These authors contributed equally to this work (x-ray, Raman, theory).

[1]Department of Physics and Frederick Seitz Materials Research Laboratory, University of Illinois, Urbana, IL 61801, USA

[2]Institute for Condensed Matter Theory, University of Illinois at Urbana-Champaign, Urbana, IL 61801, USA

[3]Cornell High Energy Synchrotron Source, Cornell University, Ithaca, NY 14850, USA

[4]Advanced Photon Source, Argonne National Laboratory, Argonne IL 60439, USA

[5]National University of Singapore, 21 Lower Kent Ridge Road, Singapore 119077

**Through its interactions with the atomic lattice and spins, the orbital degree of freedom in transition metal compounds can exhibit rich phase behavior—including orbital ordered, orbital liquid, and orbital glass phases—and can play an essential role in mediating diverse exotic phenomena, including colossal magnetoresistance (e.g., manganites), multiferroic behavior (e.g., ferrites), and even unconventional superconductivity (e.g., cuprates, ruthenates, pnictides). Unfortunately, the complex structural and magnetic transitions present in many transition metal compounds often obscure a clear view of the central role that orbitals play in mediating their rich magnetic and structural transitions. Here, we report a combined experimental and theoretical study of KCuF$_3$, which offers—because of this material's relatively simple lattice structure and valence configuration (d$^9$, i.e., one hole in the d-shell)—a particularly clear view of the essential role of the orbital degree of freedom in governing the dynamical coupling between the spin and lattice degrees of freedom. We present Raman and x-ray scattering evidence that the phase behaviour of KCuF$_3$ is dominated above the Néel temperature ($T_N$ = 40 K) by coupled orbital/lattice fluctuations that are likely associated with rotations of the CuF$_6$ octahedra, and we show that these orbital fluctuations are interrupted by a**

**static structural distortion that occurs just above $T_N$. A detailed model of the orbital and magnetic phases of $KCuF_3$ reveals that these orbital fluctuations—and the related frustration of in-plane spin-order—are associated with the presence of nearly degenerate low-energy spin-orbital states that are highly susceptible to thermal fluctuations over a wide range of temperatures. A striking implication of these results is that the ground state of $KCuF_3$ at ambient pressure lies near a quantum critical point associated with an orbital/spin liquid phase that is obscured by emergent Néel ordering of the spins; this exotic liquid phase might be accessible via pressure studies.**

Although generally regarded as a prototypical orbital ordering material, the charge-transfer insulator $KCuF_3$ has puzzling properties that defy the standard models of orbital order. For example, it has been widely assumed that $KCuF_3$ exhibits Kugel-Khomskii (KK) type orbital order [1] at temperatures $T$ below ~ 800 K,[2,3,4,5] associated with static Jahn-Teller (JT) distortions of the $CuF_6$ octahedra. However, other measurements [6,7,8] and calculations [9] suggest the presence of structural anomalies associated with the F ions in the intermediate temperature regime between the Néel temperature $T_N \approx 40$ K [3]—at which A-type (i.e., in-plane ferromagnetic, out-of-plane antiferromagnetic) spin order sets in—and the putative orbital ordering temperature of ~800 K. Furthermore, the magnetic state of $KCuF_3$ also exhibits peculiarities between $T_N$ and ~800 K that are not well understood: $KCuF_3$ exhibits strong spin fluctuations above $T_N$, as well as an extraordinarily anisotropic superexchange ratio [10] and strongly one-dimensional behaviour in the spin subsystem down to very low energy scales.[11] There have, as yet, been no explanations given for the instability in the magnetic state of $KCuF_3$ at temperatures well below 800 K, nor has it been determined whether there is a connection between the puzzling structural and magnetic instabilities at intermediate temperatures in this material.

Figure 1 presents Raman scattering evidence that $KCuF_3$ is indeed structurally unstable well below the 800 K "orbital ordering" transition. The $KCuF_3$ single crystals studied in these measurements were grown from solution by techniques described previously,[12] and consisted of > 90% volume fraction of polytype *a*, which was determined by screening the c-axis lattice parameter of the crystals using x-ray measurements. At $T$ = 10 K, the Raman spectrum is consistent with that measured previously,[13] and includes two $E_g$-symmetry phonons near 50 cm$^{-1}$ and 260 cm$^{-1}$, a $B_{1g}$-symmetry phonon near 72 cm$^{-1}$, and an $A_{1g}$-symmetry phonon near 377 cm$^{-1}$, all of which are associated with vibrations of the F-ions in the CuF$_6$ octahedra. Figure 1 shows that there are dramatic anomalies associated with the temperature dependence of some of these phonon modes in the temperature range 50 K < $T$ < 300 K. For example, the energy of the 377 cm$^{-1}$ $A_{1g}$ breathing mode increases (i.e., "hardens") with decreasing temperature (see Fig. 1, top panel), consistent with conventional anharmonic effects. However, the 50 cm$^{-1}$ $E_g$ and 72 cm$^{-1}$ $B_{1g}$ modes exhibit a roughly 10-fold decrease in linewidth (FWHM), and a 20% and 10% decrease in energy ("softening"), respectively, with decreasing temperature (Fig. 1, bottom panel), reflecting strong temperature-dependent renormalizations of these mode frequencies in the temperature range 50 K < $T$ < 300 K.

Figure 1 also shows that the frequencies of the 50 cm$^{-1}$ and 72 cm$^{-1}$ modes stabilize at temperatures below ~ 50 K, just above $T_N$, coincident in temperature with a splitting of the doubly degenerate 260 cm$^{-1}$ $E_g$ mode into two singly degenerate modes at 260 cm$^{-1}$ and 265 cm$^{-1}$ (see Fig. 1, middle panel). The 50 cm$^{-1}$ and 72 cm$^{-1}$ mode softening between 50 K and 300 K, and the splitting of the 260 cm$^{-1}$ $E_g$ mode at 50 K, offer striking evidence for structural instabilities in KCuF$_3$ at temperatures well below the putative 800 K orbital ordering temperature, which culminate in a tetragonal-to-orthorhombic (TO) transition at a temperature just above $T_N$.[14] The TO structural transition near 50 K is likely associated with rigid GdFeO$_3$-type rotations/tilts of the CuF$_6$ octahedra, as evidenced (i) by the observation that the structural fluctuations strongly renormalize the 50 cm$^{-1}$ $E_g$ and 72 cm$^{-1}$ $B_{1g}$ modes (see Fig. 1)—which are associated with rotational motions of the F-ions in the CuF$_6$ octahedra about the a/b and c axes, respectively;[13] (ii) by x-ray evidence for GdFeO$_3$-type octahedral tilts in KCuF$_3$, the observation of which is discussed below; and (iii) by the prevalence of octahedral rotations in cubic perovskites such as LaMnO$_3$, and even in isostructural and magnetically inert RbCaF$_3$,[15] which exhibits a transition involving octahedral rotations at a temperature similar to the low temperature structural transition in KCuF$_3$.

The Raman results shown in Fig. 1—in combination with the results of earlier ESR [8] and neutron scattering [11] studies—indicate several experimental facts about KCuF$_3$ that are not explained by the widely accepted models of orbital ordering: the presence of two structural transitions—near 800 K and 50 K—separated by a wide temperature regime characterized by structural and spin fluctuations; a 3D Néel ordering temperature, $T_N \approx 40$ K, that is substantially smaller than the "upper" orbital ordering temperature of ~ 800 K;[10,11] and the close coupling between structural and spin dynamics, as revealed both by the coincidence of the spin and structural fluctuations above ~50 K, and by the proximity of the structural (~50 K) (see Fig. 1) and Néel (~40 K) transitions.

To explain the puzzling features of the various orbital and magnetic phases in $KCuF_3$, we develop a model that emphasizes the primary roles of charge transfer [16] and/or cooperative JT [17] physics in determining the orbital states. Sub-dominant orbital-spin interactions then elucidate the impact of orbital physics on the development of magnetic order.[1,18] Finally, we explain the role of lattice degrees of freedom on orbital-spin interactions in stabilizing both orbital and spin order below a low-temperature structural transition in $KCuF_3$.

The Hamiltonian for the model, $H_{eff} = H_{\tau\tau} + H_{CF} + H_{\tau S} + H_{OR}$, describes a system of interacting orbital, lattice and spin degrees of freedom. The dominant term, $H_{\tau\tau} = \sum_{\langle ij \rangle} \sum_{\alpha=a,b,c} J_1^\alpha \tau_i^\alpha \tau_j^\alpha$, represents the nearest-neighbour orbital-orbital interactions. In it, the orbitals are expressed in terms of the pseudospins $\tau_i^{a/b} = (-T_i^z \pm \sqrt{3} T_i^x)/4$, $\tau_i^c = T_i^z/2$; the lattice sites are indexed by $i$ and $j$; $(T_i^x, T_i^y, T_i^z)$ are Pauli matrices acting on the subspace spanned by the degenerate pair of $e_g$ orbitals $3d_{3z^2-r^2}$ and $3d_{x^2-y^2}$; and the (true) electron spins are denoted by $\mathbf{S}_i$. As emphasized by Mostovoy and Khomskii,[16] the orbital-orbital interactions can arise either from charge-transfer processes in the presence of on-ligand Hubbard repulsion, or from the cooperative JT coupling of the orbitals to the tetragonal lattice distortions that condense at a temperature of order 800 K.[1,17] The importance of both charge-transfer and cooperative JT mechanisms in $KCuF_3$ is also supported by recent electronic structure calculations.[19] In the tetragonal phase, the condensed distortion produces a small anisotropy in the inter-orbital couplings, i.e., $(J_1^a, J_1^b, J_1^c) = (1,1,\beta)J_1$, with $\beta \approx 1$. Such a distortion can also lead to a weak tetragonal crystal-field term, $H_{CF} = \lambda_{CF} \sum_i \tau_i^z$ (with $\lambda_{CF} \ll J_1$), which would lift the orbital degeneracy. The term $H_{\tau S}$ couples the spin and orbital correlations to one another via superexchange:[1,18]

$$H_{\tau S} = \sum_{\langle ij \rangle} \sum_{\alpha=a,b,c} J_2^\alpha [4(\mathbf{S}_i \cdot \mathbf{S}_j)(\tau_i^\alpha - \tfrac{1}{2})(\tau_j^\alpha - \tfrac{1}{2}) + (\tau_i^\alpha + \tfrac{1}{2})(\tau_j^\alpha + \tfrac{1}{2}) - 1]$$
$$+ \eta \sum_{\langle ij \rangle} \sum_{\alpha=a,b,c} J_2^\alpha [(\mathbf{S}_i \cdot \mathbf{S}_j)(\tau_i^\alpha + \tau_j^\alpha - 1) + \tfrac{1}{2}(\tau_i^\alpha - \tfrac{1}{2})(\tau_j^\alpha - \tfrac{1}{2}) + \tfrac{3}{2}(\tau_i^\alpha \tau_j^\alpha - \tfrac{1}{4})]. \quad (1)$$

Here, the parameter $\eta = J_H/U \ll 1$ is the ratio of the Hund coupling $J_H$ to the Hubbard repulsion $U$ for the Cu 3d-orbitals. We shall assume that there is a large c-axis anisotropy amongst the orbital-spin couplings [i.e., $J_2^c > J_2^a (= J_2^b)$]; this contributes to a large anisotropy amongst the spin-exchange couplings. Furthermore, in keeping with experimental observations of disparate orbital and spin energy-scales in KCuF$_3$,[5] we assume that $J_1 \gg J_2^c$. Finally, through $H_{OR}$, we incorporate an additional coupling between the nearest neighbour orbital and spin correlations within the a-b planes, which are generated by orthorhombic distortions $\mathbf{Q}_{ij}$ away from the tetragonal lattice:

$$H_{OR} = -\mu \sum_{\langle ij \rangle} \sum_{\alpha=a,b} J_2^\alpha (\mathbf{S}_i \cdot \mathbf{S}_j)(\tau_i^\alpha - \tfrac{1}{2})(\tau_j^\alpha - \tfrac{1}{2}) |\mathbf{Q}_{ij}|^2. \tag{2}$$

Here, $\mathbf{Q}_{ij}$ describes fluctuations in the position of the F ligand atom transverse to the line connecting adjacent in-plane Cu sites $i$ and $j$, which our data suggest are associated with rotations of the CuF$_6$ octahedra.

To analyze this model, we take a variational approach. We begin by assuming that the crystal-field splitting $\lambda_{CF}$ is weak and that the anisotropy in the inter-orbital couplings $\{J_1^\alpha\}$ is small. We then make the hybrid orbital (HO) Ansatz for the state of a single hole with spin-projection $\sigma$ at lattice site $i$: $|i\mu\sigma\rangle = \cos(\theta_i)|iz\sigma\rangle + \sin(\theta_i)|ix\sigma\rangle$. Here, the mixing angles $\{\theta_i\}$ are our variational parameters and, motivated by the anticorrelating nature of the orbital interactions, we take them to alternate in sign between the two sublattices in the a-b plane. Next, we consider two classical spin orderings, G-type (i.e., in-plane and out-of-plane antiferromagnetic) and A-type antiferromagnetism (AFM), found prominently in KCuF$_3$ and related materials. Hence, for each of the two aforementioned spin orderings, we find that tetragonal lattice distortions lead to a low-energy orbital configuration, $|HO_1\rangle$ (see Fig. 2a) or $|HO_2\rangle$ (see Fig. 2b), each characterized by a distinct value of $\theta$. Energetically, these configurations are separated by only the smallest couplings in the problem, i.e.,

$$\Delta E = E_{HO_1} - E_{HO_2} \approx \left(\eta - \frac{1}{4}\right)J_2^{a/b} + \frac{\eta}{4} J_2^c + \frac{\mu}{16}|\mathbf{Q}|^2. \tag{3}$$

Configurationally, the mixing angle for these hybrid states lies close to $\theta = \pi/4$, while the difference between the mixing angles of the hybrid states is small and given by

$$\cos(2\theta_i^{HO_1}) - \cos(2\theta_i^{HO_2}) \approx \frac{(2-\eta)J_2^{a/b} - (\mu/8)|\mathbf{Q}|^2}{(4+2\beta)J_1}. \tag{4}$$

The effective spin-spin exchange coupling for the two orbital states is determined from the orbital-spin term using the mixing angles for these orbital states. Thus, we find that orbital state $|HO_1\rangle$ is associated with in-plane *antiferromagnetic* spin-exchange coupling

$$\Gamma_{HO1}^{a/b} \approx +\left(\tfrac{1}{4}-\eta\right)J_2^{a/b} - (\mu/8)|\mathbf{Q}|^2. \tag{5}$$

In contrast, orbital state $|HO_2\rangle$ is associated with in-plane spin-exchange coupling that is *ferromagnetic*,

$$\Gamma_{HO2}^{a/b} \approx -(7\eta/8)J_2^{a/b} - (\mu/8)|\mathbf{Q}|^2. \qquad (6)$$

However, both orbital states have the same c-axis antiferromagnetic spin-exchange coupling $\Gamma^c \approx (1-\eta)J_2^c$, which is much larger than either of the in-plane couplings. In keeping with the empirical Goodenough-Kanamori-Anderson (GKA) rules,[20] which explain low magnetic energy scales in terms of small orbital overlaps, both $|HO_1\rangle$ and $|HO_2\rangle$ are consistent with low in-plane spin-exchange couplings. In fact, the state $|KK\rangle$ of alternating $3d_{x^2-z^2}$ and $3d_{y^2-z^2}$ orbitals, originally proposed for KCuF$_3$ by Kugel and Khomskii [1] and characterized by a mixing angle $\theta = \pi/6$, is another state consistent with the GKA rules. In our model, however, the $|KK\rangle$ state is significantly higher in energy than both the $|HO_1\rangle$ and $|HO_2\rangle$ states, by an energy of order $(1+\beta/2)(J_1/4)$.

What are the implications of having nearly degenerate orbital states $|HO_1\rangle$ and $|HO_2\rangle$ as the lowest energy states in KCuF$_3$? Choosing reasonable values for the various couplings and parameters, $(J_1, J_2^c, \lambda_{CF}, J_2^{a/b}) = (600, 200, 50, 30)$ K and $(\beta, \eta) = (1.05, 0.1)$ (see Refs. [1,10,11,18,21]), we find the energy difference between $|HO_1\rangle$ and $|HO_2\rangle$ to be on the order of a few Kelvin. Consequently, throughout the temperature regime above the TO structural phase transition near 50K, thermal fluctuations readily cause both orbital states to be populated. Yet, while the two hybrid orbital states have mixing angles that differ by only a few degrees (see Fig. 2c), modest fluctuations between these two orbital states have severe consequences for in-plane magnetic ordering: because the in-plane spin-exchange couplings $\Gamma^{a/b}$ associated with states $|HO_1\rangle$ and $|HO_2\rangle$ are small but of opposite sign, even small orbital fluctuations lead to large fluctuations of the spins, disordering any long-ranged in-plane magnetic ordering.[18] This picture is consistent with the experimental observation that in-plane spin-spin correlations are very short ranged in the tetragonal phase, even down to the TO structural phase transition.[10] Furthermore, a large c-axis anisotropy in the orbital-spin couplings (i.e., $J_2^{a/b}/J_2^c \ll 1$) and a small Hund parameter ($\eta \ll 1$) together produce a large anisotropy in the spin-exchange couplings. For the parameters given above, we find the following values for the spin exchange couplings: $(\Gamma^c, \Gamma_{HO_1}^{a/b}, \Gamma_{HO_2}^{a/b}) \sim (180, 3, -2)$ K, which are consistent with known values, particularly the anisotropic spin correlations observed via neutron scattering.[5,11]

Additionally, our model can explain the anomalous phonon behaviour observed in KCuF$_3$ using Raman spectroscopy (see Fig. 1). For example, Eq. (2) shows that in-plane ferromagnetic and antiferromagnetic spin ordering have opposing effects on the stiffness associated with such lattice fluctuations: aligned neighbouring spins cause a softening of the stiffness, whilst anti-aligned neighbours cause a hardening. This can be expressed as an effective spring constant that is primarily sensitive to the spin correlations:

$$K_{eff} \approx K - 2\mu \langle \mathbf{S}_i \cdot \mathbf{S}_j \rangle \sum_{\alpha=a,b} J_2^\alpha \langle (\tau_i^\alpha - \tfrac{1}{2})(\tau_j^\alpha - \tfrac{1}{2}) \rangle. \quad (7)$$

Equation (7) confirms that the ferromagnetic (A-type) in-plane correlations that develop with decreasing temperature in KCuF$_3$ are associated with the phonon mode softening observed above $T_N$ in Fig. 1. Moreover, the observation that A-type, rather than G-type, AFM emerges at a Néel-ordering temperature below the TO structural transition shown in Fig. 1 follows from the enhancement, via lattice fluctuations, of the in-plane FM spin-exchange coupling $\Gamma_{HO2}^{a/b}$ relative to the AFM coupling $\Gamma_{HO1}^{a/b}$ [see Eqs. (5,6)].

Our model also suggests that the locking in, at temperatures below the TO transition, of orthorhombic fluctuations is associated with the emergence of bistability with respect to the positions of the in-plane F ligand atoms, each of which resides in one of a pair of locally stable off-bond sites. This bistability promotes glassiness, as it makes possible ordered domains of cooperative off-bond F-atom positions, which are separated from one another by domain walls. To test this description, we carried out a temperature dependent x-ray study of both the orbital and magnetic order in KCuF$_3$ to study the relationship between the tilting of the CuF$_6$ octahedra and the magnetic ordering transition. Soft x-ray magnetic scattering measurements were conducted at the Cu L$_{3/2}$ edge at beam line X1B at the National Synchrotron Light Source. Hard x-ray measurements were performed at Sector 4 at the Advanced Photon Source and C Line at the Cornell High Energy Synchrotron Source. We will denote momenta in terms of Miller indices, i.e., (H, K, L), indicates a momentum transfer $\mathbf{q} = (2\pi H/a, 2\pi K/b, 2\pi L/c)$, where $a = b = 5.85$ Å and $c = 7.82$ Å. Our magnetic scattering studies are summarized Fig. 3a. Below $T_N \approx 40$ K, a resolution-limited, (0,0,1) antiferromagnetic reflection is observed, in agreement with previous studies.[3,11] Furthermore, diffuse magnetic scattering is observed above $T_N$, arising from critical fluctuations in the magnetic order, as observed in Ref. [11]. The momentum linewidth is highly anisotropic, confirming the very anisotropic spin-exchange interaction $\Gamma$. This anisotropy is what gives rise to the quasi one-dimensional magnetism observed above the critical regime.[11] The correlation length in all three directions was observed to diverge simultaneously, indicating a single, three-dimensional Néel transition at 40 K.[3,10,11]

Our hard x-ray measurements are summarized in Fig. 3b, which shows scans through the (1,0,5) Bragg position as a function of temperature. The (1,0,5) position corresponds to the wave vector of the orbital order, and is symmetry-equivalent to orbital reflections studied previously.[5] Importantly, (1,0,5) is also the expected wave vector of GdFeO$_3$-type rotations of the CuF$_6$ octahedra, fluctuations of which are suggested by our Raman measurements. As in Ref. [5], the (1,0,5) reflection is present and sharp at room temperature (≈300 K), consistent with some degree of orbital ordering in KCuF$_3$. Below 300 K, however, there is an observed enhancement of scattering at (1,0,5)—reaching a maximum intensity at $T$~100 K, i.e., in the middle of the fluctuational regime observed with Raman (see Fig. 1)—which is consistent with the presence of dynamical, GdFeO$_3$-type rotations of the CuF$_6$ octahedra in this temperature range. Note that because these rotations displace the apical F atoms, they give rise to a dynamical Dzyaloshinsky-Moriya interaction, as observed by several groups.[8,23] Finally, below the temperature (~50 K) at which the 260 cm$^{-1}$ E$_g$-symmetry Raman mode splits (see Fig. 1), the (1,0,5) reflection drops in intensity and acquires a series of diffuse sidebands, indicating that the GdFeO$_3$-type rotations become static and disordered below ~50 K. These sidebands exhibit hysteresis: with subsequent warming and cooling through the transition, the diffuse scattering is not reversible, suggesting a glassy nature to GdFeO$_3$-type order.

What are the lessons learned from the results presented here? Orbital ordering in KCuF$_3$ occurs not via a single transition near 800 K, as is commonly believed, but rather in two stages: the higher temperature transition reduces the manifold of well-populated orbital states, but leaves a pair of nearly degenerate (to within a few Kelvin) hybrid states that have similar orbital configurations, but very different in-plane spin correlations. The near degeneracy of these orbital states accounts for the large temperature range over which structural and in-plane spin fluctuations are observed, and for the large anisotropy in spin correlation lengths that have been measured, in KCuF$_3$ above $T_N$. Our x-ray data suggest that these structural fluctuations are associated with CuF$_6$ octahedral rotations, the presence of which accounts for the dynamical Dzyaloshinsky-Moriya interaction observed previously in KCuF$_3$.[8,23] A second structural transition near 50 K in KCuF$_3$—observed via Raman scattering and associated with a freezing in of the CuF$_6$ octahedral rotations/tilts—is necessary to stabilize the lowest energy orbital state, i.e., that associated with A-type antiferromagnetic order, accounting for the dramatically different energy scales associated with $T_N$ (~40 K) and the structural transition near 800 K in KCuF$_3$. Finally, the strong coupling between the lattice and spin degrees of freedom—mediated through the orbitals—accounts for both the anomalous softening of certain Raman modes in KCuF$_3$ and the strong correlation between structural and magnetic fluctuations/transitions in this material.

The results presented here have several important implications, all of which point to KCuF$_3$ as an ideal system for studies of orbital physics. First, the strong, orbitally mediated coupling of spin- and lattice-degrees of freedom in KCuF$_3$ suggests that this material is ideal for studying how orbitals influence both magnetic and elastic properties in response to applied magnetic field and pressure. Second, we note that KCuF$_3$ is uniquely suited for the study of orbital fluctuations and their impact on spin- and lattice dynamics—in that the large separation between structural transitions in this material provides an especially broad temperature range over which orbital fluctuations can be observed. Nevertheless, the physics described by our model implies that orbital fluctuations should influence the behaviour of other orbital ordering materials, such as LaMnO$_3$ and LiNiO$_3$, and thus it is important to look for similar fluctuational behaviour in these materials. Finally, our results suggest that the ground state of KCuF$_3$ is close to an orbital/spin liquid phase—associated with fluctuations between states $|HO_1\rangle$ and $|HO_2\rangle$ (see Fig. 2)—which is pre-empted by a low-temperature structural phase transition that stabilizes A-type antiferromagnetic order. Indeed, a quantum phase transition to an orbital/spin liquid state may be achievable in KCuF$_3$ with increasing pressure, which favors more isotropic structural configurations,[23] and hence should frustrate the low temperature structural transition and reduce the splitting between orbital states $|HO_1\rangle$ and $|HO_2\rangle$.

**Acknowledgments** We gratefully acknowledge helpful input from D. I. Khomskii, M.V. Mostovoy, A.J. Millis, A.K. Sood, and H.R. Krishnamurthy. This work was supported by the U.S. Department of Energy, Division of Materials Sciences through the Frederick Seitz Materials Research Laboratory at the University of Illinois at Urbana-Champaign under Award No. DE-FG02-07ER46453, with soft x-ray studies supported by Award No. DE-FG02-06ER46285. Use of the Advanced Photon Source was supported by contract # DE-AC02-06CH11357. Use of the NSLS was supported by contract # DE-AC02-98CH10886. Use of CHESS was supported by the National Science Foundation under award DMR-0225180.

**Competing interests statement** The authors declare that they have no competing financial interests.

**Correspondence** and requests for materials should be addressed to P.A. (abbamonte@mrl.illinois.edu) or S.L.C. (slcooper@illinois.edu).

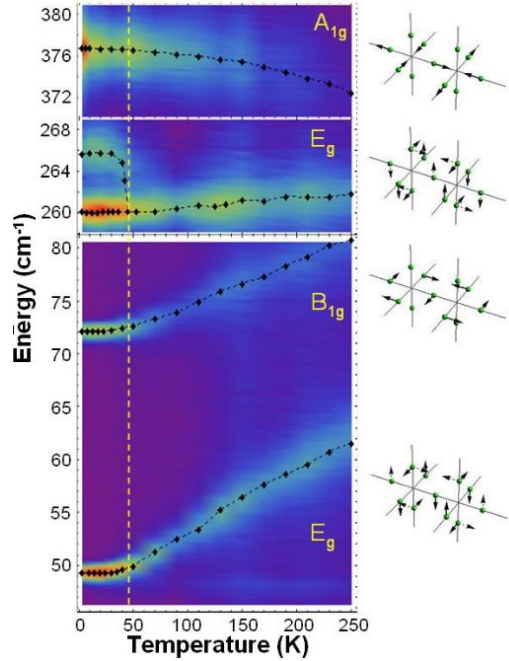

**Figure 1** (left) Temperature dependence of select Raman-active phonons in $KCuF_3$, showing a dramatic softening and narrowing of the lowest energy $E_g$ and $B_{1g}$ modes (bottom panel) with decreasing temperature—indicative of a structurally fluctuating regime above 50 K—and a splitting of the higher energy $E_g$ mode below 50 K, consistent with a static orthorhombic distortion assocated with rotations/tilts of the $CaF_6$ octahedra. (right) Eigenvectors associated with the various phonon modes. Redder regions in the contour plot indicate high phonon scattering intensities and bluer regions indicate low phonon scattering intensities.

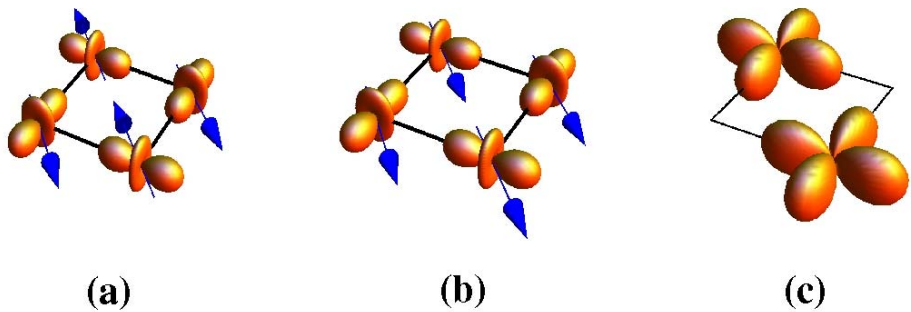

**Figure 2** Nearly degenerate hybrid orbital states (a) $|HO_1\rangle$ and (b) $|HO_2\rangle$, which are both thermally occupied in the intermediate temperature regime. The second hybrid pattern is stabilized by the low-temperature orthorhombic distortion. Blue arrows indicate spin direction on the Cu site. (c) Orbital fluctuations between the two hybrid orbital states.

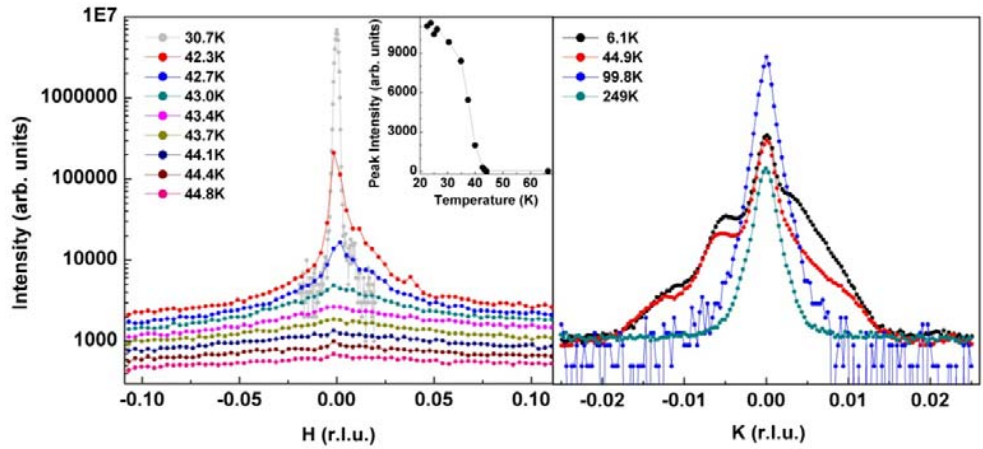

**Figure 3**. Structural and magnetic x-ray data of $KCuF_3$. (a) Resonant scattering near the Cu $L_{3/2}$ edge, showing critical magnetic fluctuations above $T_N$ around the (0,0,1) antiferromagnetic reflection, which are contrasted with the resolution-limited peak at 30.7 K, below $T_N$. (inset) Temperature dependence of the magnetic order parameter around the (0,0,1) reflection. (b) Non-resonant scattering at 8.8 keV, showing the temperature dependence of the (1,0,5) Bragg reflection, which serves as an order parameter for $GdFeO_3$-type distortions of the $CuF_6$ octahedra.